\documentclass[letterpaper]{article} 
\usepackage{aaai2026}  
\usepackage{times}  
\usepackage{helvet}  
\usepackage{courier}  
\usepackage[hyphens]{url}  
\usepackage{graphicx} 
\usepackage{natbib}  
\usepackage{caption} 
\usepackage{amsmath}
\usepackage{amssymb}
\usepackage{bm}
\usepackage{booktabs}
\usepackage{makecell}
\usepackage{multirow}
\newcounter{mycounter1}
\newcounter{mycounter2}
\title{PS-GS: Gaussian Splatting for Multi-View Photometric Stereo}
\author{
    Yixiao Chen\textsuperscript{\rm 1},
    Bin Liang\textsuperscript{\rm 2},
    Hanzhi Guo\textsuperscript{\rm 1},
    Yongqing Cheng\textsuperscript{\rm 1},
    Jiayi Zhao\textsuperscript{\rm 1},
    Dongdong Weng\textsuperscript{\rm 1,}\textsuperscript{\rm 3}\thanks{Corresponding author.}
}
\affiliations{
    \textsuperscript{\rm 1}Beijing Engineering Research Center of Mixed Reality and Advanced Display, School of Optics and Photonics, Beijing Institute of Technology, China\\
    \textsuperscript{\rm 2}China Software Testing Center, Beijing, China\\
    \textsuperscript{\rm 3}Zhengzhou Research Institute, Beijing Institute of Technology, China\\

    crgj@bit.edu.cn
}

\begin{document}

\maketitle

\begin{abstract}
Integrating inverse rendering with multi-view photometric stereo (MVPS) yields more accurate 3D reconstructions than the inverse rendering approaches that rely on fixed environment illumination. However, efficient inverse rendering with MVPS remains challenging. To fill this gap, we introduce the Gaussian Splatting for Multi-view Photometric Stereo (PS-GS), which efficiently and jointly estimates the geometry, materials, and lighting of the object that is illuminated by diverse directional lights (multi-light). Our method first reconstructs a standard 2D Gaussian splatting model as the initial geometry. Based on the initialization model, it then proceeds with the deferred inverse rendering by the full rendering equation containing a lighting-computing multi-layer perceptron. During the whole optimization, we regularize the rendered normal maps by the uncalibrated photometric stereo estimated normals. We also propose the 2D Gaussian ray-tracing for single directional light to refine the incident lighting. The regularizations and the use of multi-view and multi-light images mitigate the ill-posed problem of inverse rendering. After optimization, the reconstructed object can be used for novel-view synthesis, relighting, and material and shape editing. Experiments on both synthetic and real datasets demonstrate that our method outperforms prior works in terms of reconstruction accuracy and computational efficiency.
\end{abstract}

\section{Introduction}
Inverse rendering (IR) focuses on reconstructing geometry, materials, and lighting from captured posed images. Based on IR, many downstream tasks such as relighting and material editing can be implemented, which play a pivotal role in various fields, such as games, augmented reality, cultural heritage, and film production. However, IR is still a long-term problem in computer graphics and vision as it is an ill-posed problem, particularly when input images are captured in an unknown environment illumination or with sparse views. The success of neural radiance field (NeRF) \cite{nerf}, which leverages a directional-aware multi-layer perceptron (MLP) to represent 3D scenes through volume rendering, has inspired many NeRF-based approaches \cite{neilf, neilf++, pbrnerf} to proceed with the IR and address the ill-posed issue. Although these methods demonstrate compelling reconstruction of geometry and reflectance, they still face challenges in terms of explicit editing and excessive computational costs, which limit their further application. 

More recently, 3D Gaussian Splatting(3DGS) \cite{3dgs} has attracted much attention due to its ability of real-time and high-fidelity rendering in novel view synthesis (NVS). 3DGS models a 3D scene by a set of explicit 3D Gaussian primitives and renders pixels by alpha-blending through a specially designed rasterizer, which can be exploited to achieve efficient inverse rendering. However, as its difficulty in accurately simulating ray-based effects, some 3DGS-based IR methods have adopted simplified versions of the rendering equation \cite{gsshader,gs-ir} or proposed a 3D Gaussian ray-tracing technique \cite{r3dg}. Another challenge of these methods is the struggle with accurately computing normal only by a depth-related regularization without exploiting the relationship between primitive normal and geometry attributes. To this end, 2D Gaussian Splatting (2DGS) \cite{2dgs} and 2DGS-based works \cite{irgs, ref-gs} leverage the normal of the elliptical disk as the normal for each primitive, which acquires more compelling normal results than 3DGS-based methods. However, as the lack of additional information in optimization, the rendered normal maps are still over-smooth and lack detail. Moreover, when proceeding with sparse-view reconstruction, the ill-posed problem gets worse, which leads to a decline in the overall quality of reconstruction.

On the other hand, photometric stereo (PS) is a technique that utilizes captured single-view images illuminated by diverse directional lights (multi-light) to obtain per-pixel surface normal of the scene. Although their ability to recover fine surface details, PS approaches \cite{sdps, unips} with single-view input are not capable of reconstructing a full 3D shape. To simultaneously obtain surface details and full 3D shape, the multi-view photometric stereo (MVPS) \cite{mvps1, mvps2} method that combines multi-view stereo (MVS) method with PS is proposed . By incorporating the neural implicit field and the point-based splatting, PS-NeRF \cite{ps-nerf} and DPIR \cite{dpir} provide the solution to MVPS-based inverse rendering, respectively. Despite their advancements, they either suffer from excessive computational costs like congeners or use a simplified rendering equation that is unable to acquire more diverse materials.

To address the aforementioned issues altogether, we propose Gaussian Splatting for Multi-view Photometric Stereo (\textbf{PS-GS}), an inverse rendering approach that extends 2D Gaussian Splatting to physically-based deferred rendering with MVPS. We adopt the strategy of PS-NeRF \cite{ps-nerf} for optimization, which leverages guidance normals estimated via uncalibrated photometric stereo to regularize the rendered normal maps and uses multi-view and multi-light images for training. Benefiting from the efficient 2D Gaussian ray-tracing technique proposed by IRGS \cite{irgs}, we modify this technique to estimate the visibility for the object illuminated under single directional light (SDL), and leverage its visibility result to regularize the lighting.

Our method starts with pretraining a standard 2D Gaussian splatting model. Based on this model, we then use the full physically-based rendering equation without simplification to jointly optimize materials, geometry, and lighting. In the rendering equation, a multi-layer perceptron (MLP) is leveraged to model lighting that is not modeled by the parameters of Gaussian primitives. The rendered normal maps and incident lighting are regularized by the UPS-estimated normals and SDL 2D Gaussian ray-tracing computed visibility, respectively. We exploit pre-acquisition of diverse parameter maps as regularization to mitigate the ill-posed problem of inverse rendering. To efficiently render the multi-view and multi-light images, we proceed with the shading in image space, which can both achieve rapid rendering and improve reconstruction quality. 

To the best of our knowledge, this is the first method to integrate Gaussian splatting and MVPS, which can accurately and efficiently proceed with geometry, materials, and lighting estimation and enable novel view synthesis, relighting, and shape and material editing. Experiments on both real and synthetic dataset confirm that our method has improved training time and reduced the usage of GPU memory compared to previous MVPS-based inverse rendering approaches while achieving comparable or better results. 
In summary, contributions of this paper are as follows:
\begin{itemize}
\item We innovate an efficient physically-based deferred inverse rendering approach for multi-view photometric stereo, which jointly optimizes geometry, materials, and lighting based on 2DGS.
\item We propose to refine the rendered normal maps of Gaussian model by the UPS estimated normals, which significantly enhances the quality of surface estimation.
\item We modify 2D Gaussian ray-tracing to suit SDL and leverage its visibility results to regularize the MLP-computed lighting, which improves the lighting reconstruction.
\item Our method achieves compelling results on many tasks, even for sparse-view inputs, including NVS, relighting, and material and shape editing.
\end{itemize}

\section{Related work}
\subsubsection{Novel view synthesis (NVS)} is a long-standing problem in computer vision and graphics, which focuses on generating images under unseen viewpoints by a collection of captured images of the scene. Neural radiance fields (NeRF) \cite{nerf} leverage neural networks to represent the scene as a continuous 5D function, which has catalyzed a trend in high-fidelity NVS of complicated 3D scenes. Subsequent developments have improved the NeRF in accelerating training speed \cite{instant-ngp, efficient-nerf}, enhancing the rendering quality \cite{mip-nerf, trip-nerf}, enabling sparse views reconstruction \cite{sparse-nerf, zerorf}, etc. More recently, 3D Gaussian Splatting (3DGS) \cite{3dgs} renders explicit 3D Gaussian opacity fields by a well-designed tile-based rasterization pipeline, achieving the state-of-the-art performance on high-fidelity real-time NVS. Many following works have been inspired by 3DGS to achieve various tasks, such as dynamic scene reconstruction \cite{defgs, stgs}, animatable avatar \cite{ani-ava,egha}, and SLAM \cite{slam-gs, gs-slam}. Based on 3DGS, 2D Gaussian Splatting (2DGS) \cite{2dgs} has further expanded geometry reconstruction, which also improves the ability to simulate ray-based effects. In this paper, we adopt 2D Gaussian primitives to represent scenes and proceed with SDL ray-tracing on them, which can lead to more accurate surface and lighting reconstruction.

\subsubsection{Inverse rendering (IR)} aims to reconstruct materials and lighting from observed images, which is modeled as the interaction of the illumination and surface geometry. However, it suffers from the ill-posed problem due to the inherent ambiguity between captured images and underlying properties. Inspired by the success of NeRF in 3D representation, NeRF-based IR methods model the interaction of light with the neural sense representation that contains various material properties \cite{nerfactor, irgo}. However, all these approaches almost suffer from long training times. Recent methods have combined the 3DGS with IR by attaching lighting and material parameters to each Gaussian primitive. Owing to its specially designed rendering pipeline, 3DGS-based approaches significantly outperform the NeRF-based works on reconstruction quality and training speed. GS-IR \cite{gs-ir} realizes 3DGS-based IR by simplifying the rendering equation via split-sum approximation and using baked volumes to tackle the occlusion in modeling indirect illumination. R3DG \cite{r3dg} renders each Gaussian individually with full rendering equation, using 3D Gaussian ray-tracing to compute visibility. GShader \cite{gsshader} applies an explicit approximation to the rendering equation by simplified shading functions and predicts normal residual for precise normal estimation. However, methods based on 3DGS compute normals solely from the depth information of Gaussian primitives, without leveraging other geometric cues, thus facing challenges in achieving high-precision surface estimation. The 2DGS-based approaches, whose normal of each Gaussian corresponds directly to the surfel normal, demonstrate strong performance in both accurate normal estimation and shape reconstruction by integrating depth information. Ref-Gaussian \cite{ref-gs}, built upon 2DGS, models reflective objects by leveraging pre-integration to simplify the rendering equation and computing the visibility of reflection terms via ray-tracing based on the extracted mesh. IRGS \cite{irgs} proposes a 2D Gaussian ray-tracing technique to model inter-reflections without simplifying the rendering equation for inverse rendering. However, as constant environment illumination of these works, the inherent ambiguity remains, leading to over-smooth surface reconstruction.

\subsubsection{Multi-view photometric stereo (MVPS)} leverages multi-view captured images under varying illumination to realize more accurate 3D surface reconstruction. Traditional approaches \cite{mvps1, mvps2} leverage simplified surface reflectance and separately reconstruct a coarse shape by MVS and per-view surface normal by PS, and the obtained normal is then used to refine the coarse shape. More recently, MVPS has been combined with implicit neural representation or point-based model to acquire high-fidelity reconstruction results. \citeauthor{kaya} first proposed the NeRF-based MVPS approach that integrates depth maps with normal maps from a pretrained PS network to reconstruct complete shape. PS-NeRF \cite{ps-nerf} jointly estimates the geometry, material, and illumination by NeRF-based shadow-aware inverse rendering, and proceeds with regularization on surface normal by the normals estimated from UPS method \cite{sdps}. RNb-Neus \cite{rnb-neus} leverages both explicit 2.5D shape representation and implicit neural shape representation to accurately reconstruct the shape. DPIR \cite{dpir} exploits hybrid point-volumetric geometry representation and specially designed point-based visibility detection method to jointly optimize the point locations, radii, surface normals, and reflectance with differentiable point-based rendering. However, these methods either suffer from excessive training time, are merely applicable to shape reconstruction, or are incapable of recovering materials of the surface.

\section{Preliminary}
\subsection{2D Gaussian Splatting}
In standard 2DGS \cite{2dgs}, each Gaussian is represented as 2D oriented planar disk, which is defined by a position $\bm{p}$, two scaling factors $s_u$ and $s_v$, two principal tangential vector \bm{$t_u$} and \bm{$t_v$}, opacity $o$, and color coefficients $\bm{c}$. The influence of each 2D Gaussian is defined as:
\begin{equation}
\mathcal{G}(\mathbf{u}(\bm{x})) = \exp\left(-\frac{u^2 + v^2}{2}\right),
\label{eq1}
\end{equation}
where $(u, v)$ are ray-splat intersection coordinates in the local tangent space, $\bm{x}$ represents the homogeneous ray passing through pixel $(x, y)$ and intersecting the 2D Gaussian primitive at $z$. $\mathbf{u}(\bm{x})=(u, v)$ is the intersection point function, which can be described by:
\begin{equation}
\bm{x}=(xz,yz,z,1)^{T}=\bm{W}\bm{H}(u,v,1,1)^{T},
\label{eq2}
\end{equation}
where $\bm{W}$ is the transformation matrix from world space to camera space and
\begin{equation}
\bm{H}=\begin{bmatrix}
  s_{u}\bm{t_{u}} & s_{v}\bm{t_{v}} & \bm{0} & \bm{p} \\
  0&  0&  0& 1
\end{bmatrix}
\in \mathbb{R}^{4\times 4}.
\label{eq3}
\end{equation}

Unlike 3DGS, which struggles to directly derive normals from geometry parameters, the normal for each 2D Gaussian is defined as the cross product of two tangential vectors of the Gaussian disk:
\begin{equation}
\bm{n}=\bm{t_{u}}\times \bm{t_{v}}. 
\label{eq4}
\end{equation}

During rendering, 2D Gaussians are projected in camera space. The final rendered pixel color $\bm{C}$ is calculated by:
\begin{equation}
\bm{C}_i = \sum_{n \in N} \bm{c}_n \alpha_n T_n, \quad T_n = \prod_{m=1}^{n-1} (1 - \alpha_m),
\label{eq5}
\end{equation}
where $T_n$ is the accumulated transparency and $\alpha_n=o_n\cdot\mathcal{G}_n(\mathbf{u}(\bm{x}))$. See more details of 2DGS in \cite{2dgs}.

\subsection{Rendering equation} 
In PS-GS, we replace spherical harmonics (SH) lighting with physically-based rendering (PBR) and adopt deferred shading \cite{deferred-shading}. The illumination at a point $\bm{r}$ is given by the rendering equation:
\begin{equation}
L(\bm{\omega}_o, \bm{r}) = \int_{\Omega} f(\bm{\omega}_o, \bm{\omega_i}, \bm{r}) L_i(\bm{\omega}_i, \bm{r}) (\bm{\omega}_i \cdot \bm{n}) \, d\bm{\omega}_i,
\label{eq6}
\end{equation}
where $\bm{\omega}_o$ is the outgoing radiance direction, $\bm{\omega}_i$ is the incident radiance direction, $\bm{n}$ is the surface normal of point $\bm{r}$, $f$ is the bidirectional reflectance distribution function (BRDF), and $L_i (\bm{\omega}_i,\bm{r})$ is the incident radiance. The simplified Disney BRDF model \cite{disney-brdf} is used in this work, which contains only diffuse albedo $\bm{\alpha}$, metallic $m$, and roughness $r$ as parameters. Each parameter is the extra property of the Gaussian primitive. The BRDF in Equation \ref{eq6} can be devided into diffuse term $f_d$ and a specular term $f_s$ as follow:
\begin{equation}
{f_d} = \frac{1-m}{\pi}\bm{\alpha}, \quad {f_s} = \frac{DFG}{4(\bm{\omega}_i \cdot \bm{n})(\omega_o \cdot \mathbf{n})},
\label{eq7}
\end{equation}
where $D$, $F$, and $G$ represent the $GGX$ normal distribution function, the Fresnel term, and the geometry term, respectively. See more details of the rednering equation that is uesd in this work in supplementary materials.

\section{Gaussian Splatting for Multi-view Photometric Stereo}
\begin{figure*}[t]
\centering
\includegraphics[width=1\textwidth]{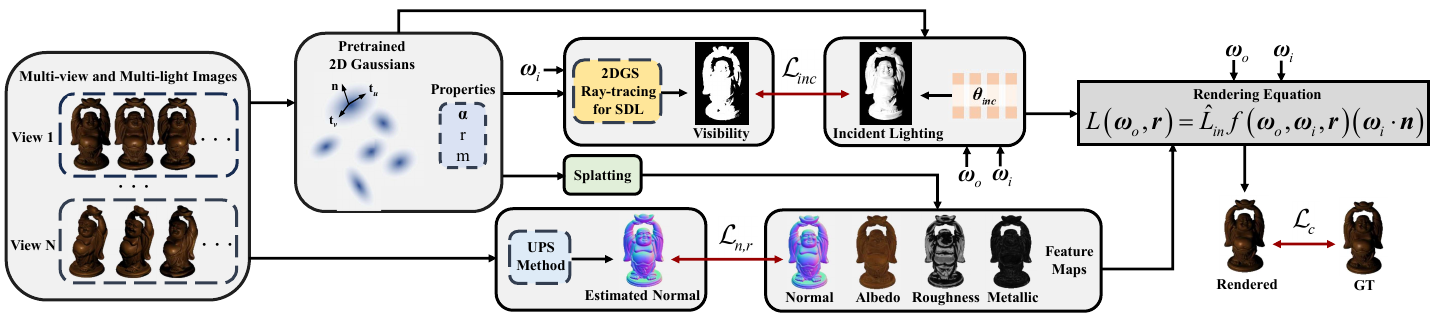} 
\caption{Overview of \textbf{PS-GS}. Starting with a set of multi-view and multi-light images, we pretrain a standard 2D Gaussian model for geometry initialization. Based on the pretrained model, we jointly optimize the geometry, materials, and lighting. To encourage accurate surface reconstruction, we compute guidance normals by the uncalibrated photometric stereo (UPS) method to regularize the rendered normal maps from the Gaussian model. We also modify the 2D Gaussian ray-tracing technique to suit single directional light (SDL), and leverage its visibility results to refine the incident lighting.}
\label{fig1}
\end{figure*}
Our goal is to simultaneously estimate geometry, materials, and lighting by multi-light images of a 3D object captured from N views. We define the multi-light images of each view $m$ as $I^{m}=\{I_{1}^{m},I_{2}^{m},\dots,I_{l}^{m}\}$, where $l$ denotes the index of light. Notably, we only focus on object-level IR in this work, and scene-level IR is beyond the scope.

Figure \ref{fig1} illustrates the overall pipeline of our method. The proposed PS-GS consists of two stages to achieve efficient Gaussian-based inverse rendering for MVPS. In the first stage, we pretrain a standard 2DGS model as initialization. In the second stage, based on a pretrained model, we jointly optimize the surface normals, materials, and lighting, using the full rendering equation with a lighting prediction network. We also leverage multi-light images to estimate the guidance normals of the object for each view by UPS method, and the normal maps of the Gaussian model in both stages are regularized by the guidance normals. Moreover, we use the visibility computed by the modified 2D Gaussian ray-tracing that is suitable for SDL to regularize the MLP-computed incident lighting.

\subsection{Stage \Roman{mycounter1}: 2D Gaussian Pretraining}
Before estimating geometry, materials, and lighting by inverse rendering with MVPS, we first train a standard 2DGS model. This stage can significantly decrease the training time by rapidly providing a reliable initialized geometry for the following inverse rendering process. During the rendering process, we also render depth ${\mathcal{D}}$ and normal maps ${\mathcal{N}}$, which is described by:
\begin{equation}
\{{\mathcal{D}}, {\mathcal{N}}\} = \sum_{i=1}^{N} w_i \{\bm{d}_i, \bm{n}_i\}, \text{ where } w_i = \frac{T_i a_i}{\sum_{i=1}^{N} T_i a_i}.
\label{eq8}
\end{equation}

In Stage \Roman{mycounter1}, We optimize our model from a set of posed multi-view light-averaged images and their masks. The total loss function of this stage is defined as:
\begin{equation}
\mathcal{L}_{\text{Stage \Roman{mycounter1}}} = \mathcal{L}_c + \lambda_{n,c} \mathcal{L}_{n,c} + \lambda_{n,r} L_{n,r} + \lambda_o \mathcal{L}_o,
\label{eq9}
\end{equation}
where $\mathcal{L}_c$ and $\mathcal{L}_{n,c}$ are the RGB reconstruction loss and normal consistency loss from seminal 2DGS \cite{2dgs}, respectively. $\lambda$ denotes the corresponding loss weight.

To enhance the shape reconstruction and surface details, we leverage the normals estimated by the UPS method using multi-light images to regularize the rendered normal map:
\begin{equation}
\mathcal{L}_{n,r} = \sum_{i=1}^{P} \left(\mathcal{N}_r - T_{c2w}(\mathcal{N}_e) \right),
\label{eq10}
\end{equation}
where $\mathcal{N}_r$ is the rendered normal map from Gaussian model, $\mathcal{N}_e$ is the UPS estimated normal, and $T_{c2w}$ is the transformation from camera coordinate system to the world coordinate system. To decrease the creation of Gaussians outside the mask in image space, we also use a binary cross mask entropy loss:
\begin{equation}
\mathcal{L}_o = -M \log O - (1 - M) \log (1 - O).
\label{eq11}
\end{equation}

\subsection{Stage \Roman{mycounter2}: Inverse Rendering with MVPS}
With the initial geometry parameters (i.e., $\bm{p}$, $\bm{t_u}$, $\bm{t_v}$, $s_u$, $s_v$, $o$, $\bm{n}$) from Stage \Roman{mycounter1}, based on multi-view and multi-light images, we perform joint optimization using the full rendering equation with MLP-computed incident lighting. Different from previous works, such as GS-IR \cite{gs-ir}, we utilize deferred shading to capture sharper highlights and decrease the GPU memory usage. In the following subsections, we will describe each component of our method in detail. 
\subsubsection{Shape and Material Modeling} 
Based on the pretrained 2D Gaussian model as initialized shape, the full rendering equation with BRDF model is used. We set the BRDF materials (i.e., $\bm{\alpha}$, $r$, and $m$) are the learnable parameters of each Gaussian, which is similar to the position $\bm{p}$, and obtain the per-Gaussian normal by Equation \ref{eq4}. Then, the PBR with deferred shading is performed based on pixel-level feature maps given by a 2D Gaussian alpha-blending process:
\begin{equation}
\bm{Q} = \sum_{n \in N} \bm{q}_n \alpha_n T_n \text{, where } \bm{Q} = \begin{bmatrix} \bm{A} \\ M \\ R \\ \bm{N} \end{bmatrix}, \, \bm{q}_i = \begin{bmatrix} \bm{\alpha}_i \\ m_i \\ r_i \\ \bm{n}_i \end{bmatrix}.
\label{eq12}
\end{equation}

Gaussian splatting based deferred shading treats alpha-blending as a smoothing filter, enabling stabilized material optimization and yielding more reliable reconstruction. To further refine the surface details, we apply the regularization between the UPS estimated normals and rendered normals like Stage \Roman{mycounter1}.
\subsubsection{Light Modeling} Explicit parameterizing all single directional incident lights as learnable parameters of each Gaussian to be rendered is expensive. Rather than modeling the incident lighting explicitly, we utilize a global neural network to estimate all the single directional lighting for each Gaussian. This implicit incidence lighting network is described by:
\begin{equation}
\hat{L}_{in} = f_{\theta} \big(\bm{p}, \bm{t}, \bm{s}, \bm{n}, \bm{\omega}_{in}, \bm{\omega}_{out} \big),
\label{eq13}
\end{equation}
where $f_\theta$ and $\theta$ are the neural network and its parameters, respectively. We also apply a Fourier encoding to position $\bm{p}$ as in NeRF \cite{nerf}, which is omitted in Equation \ref{eq13}. By exploiting the neural network to predict the incident lighting, we let the network implicitly learn the local and global lighting in the scene, which enables a GPU memory-efficient lighting modeling.

In order to constrain the neural network to predict physically reasonable results in radiance intensity and shadow, we modified the 2D Gaussian ray-tracing technique from IRGS \cite{irgs} to be suitable for SDL setup. Specifically, we replace Monte Carlo sampling by SDL sampling in 2D Gaussian ray-tracing and perform ray-tracing on the reconstructed Gaussian model. To make training efficient, ray-tracing is only applied on the pretrained model obtained in Stage \Roman{mycounter1}, as the object's geometry exhibits slight changes in Stage \Roman{mycounter2}. Finally, 
we perform $\mathcal{L}_{inc}$, which is a $\mathcal{L}_1$ loss, on the ray-tracing computed visibility and the network outputted lighting.
\subsubsection{Rendering} In Stage \Roman{mycounter2}, each image is illuminated by a specific directional light, and the rendering equation at this condition can be rewritten as:
\begin{equation}
L(\bm{\omega}_o, \bm{r}) = \hat{L}_{in} f(\bm{\omega}_o, \bm{\omega}_i, \bm{r}) (\bm{\omega}_i \cdot \bm{n}),
\label{eq14}
\end{equation}
where surface positions $\bm{r}$ can be derived from the rendered depth and normal maps for each pixel coordinate. 
\subsubsection{Total Training Loss} The total training loss function utilized for this stage is:
\begin{equation}
\mathcal{L}_{\text{Stage \Roman{mycounter2}}} = \mathcal{L}_c + \lambda_{n,c} \mathcal{L}_{n,c} + \lambda_{n,r} \mathcal{L}_{n,r} + \lambda_o \mathcal{L}_o + \lambda_{inc} \mathcal{L}_{inc},
\label{eq15}
\end{equation}
which combines the loss function of Stage \Roman{mycounter1} and the incident lighting regularization term. $\lambda$ denotes the corresponding loss weight.

\section{Experimental}
PS-GS enables efficient and accurate reconstruction of geometry, materials, and lighting. In the following, we present the comparison and ablation experiment results of PS-GS, which is evaluated on synthetic and real-world datasets. Please refer to supplementary materials for implementation details.

\subsection{Datasets and metrics}
As MVPS is uesd in our method, we conduct experiments on multi-view multi-light datasets for the inverse rendering task, including a synthetic dataset: PS-NeRF Synthesis dataset \cite{ps-nerf} (including the object named Armadillo and Bunny), and a real-world dataset: DiLiGenT-MV dataset \cite{diligent-mv} (including the object named Bear, Buddha, Reading, Cow, and Pot2). We adopt commonly used quantitative metrics for different results. Specifically, we use PSNR, SSIM \cite{ssim}, and LPIPS \cite{lpips} to evaluate the reconstructed NVS and relighted images. The mean angular error (MAE) in degrees is used for surface normal evaluation under test views. We compare PS-GS to two state-of-the-art MVPS-based inverse rendering methods: PS-NeRF and DPIR \cite{dpir}, as well as two state-of-the-art 2DGS-based inverse rendering methods: R3DG \cite{r3dg} and IRGS \cite{irgs}. 

For the comparison experiments, we have two configurations. As R3DG and IRGS are assumed to have constant environment illumination, we leverage the lighting-averaged images similar to Stage \Roman{mycounter1} to train on 15 views and test on 5 novel views. Since PS-NeRF, DPIR, and our PS-GS render multi-view and multi-light images, similar to DPIR, we use 15 views and 16 lightings for training and use 5 novel views and 96 lightings for testing. For the comparison with R3DG and IRGS, we only compute the MAE of rendered normal maps, as the rendered images of these works vary from the MVPS-based works. This comparison strategy is adopted from the DPIR and PS-NeRF. For the comparison with DPIR and PS-NeRF, we simultaneously evaluate NVS, relighting, and normal MAE without any averaging. 

\begin{table}[ht]
\setlength{\tabcolsep}{1mm}
\small
\centering
\begin{tabular}{lccccccc}
\toprule
     & Bear & Buddha & \makecell{Read-\\ing} & Cow & Pot2 & \makecell{Arma-\\dillo} & Bunny\\
\midrule
 &\multicolumn{7}{c}{PSNR$\uparrow$}\\
\midrule
\multicolumn{1}{l|}{PS-NeRF} & 34.72 & 31.69 & 32.85 & 37.17 & 39.74 & 31.13 & 32.44\\
\multicolumn{1}{l|}{DPIR} & \textbf{40.22} & 33.98 & 33.21 & 39.30 & 36.76 & 30.11 & 31.33\\
\multicolumn{1}{l|}{Ours} & 39.95 & \textbf{37.54} & \textbf{34.63} & \textbf{41.76} & \textbf{41.62} & \textbf{33.32} & \textbf{32.87}\\
\midrule
&\multicolumn{7}{c}{SSIM$\uparrow$}\\
\midrule
\multicolumn{1}{l|}{PS-NeRF}& 0.982 & 0.962 & 0.972 & 0.989 & 0.986 & 0.980 & 0.983\\
\multicolumn{1}{l|}{DPIR} & 0.981 & 0.966 & 0.980 & 0.990 & 0.983 & 0.978 & 0.980\\
\multicolumn{1}{l|}{Ours}&\textbf{0.984} & \textbf{0.977} & \textbf{0.983} & \textbf{0.992} & \textbf{0.989} & \textbf{0.987} & \textbf{0.987}\\
\midrule
 &\multicolumn{7}{c}{LPIPS$\downarrow$($\times 10^{-2}$)}\\
\cmidrule{1-8}
\multicolumn{1}{l|}{PS-NeRF} & 3.91 & 3.62 & 2.86 & 2.22 & 2.96 & 2.08 & 1.50 \\
\multicolumn{1}{l|}{DPIR} & 2.32 & 2.71 & 1.85 & \textbf{0.642} & 1.28 & 3.93 & 2.40 \\
\multicolumn{1}{l|}{Ours} & \textbf{2.08} & \textbf{1.72} & \textbf{1.40} & 0.931 & \textbf{0.927} & \textbf{1.83} & \textbf{1.21}\\
\cmidrule{1-8}
\end{tabular}
\caption{Quantitative comparison of novel-view synthesis and relighting on the objects of DiLiGenT-MV dataset and PS-NeRF synthesis dataset. In this table, the best results are in bold.}
\label{table1}
\end{table}
\subsection{Comparation Results} 
\subsubsection{Results for NVS and Relighting} 
\begin{figure}[htb]
\centering
\includegraphics[width=0.47\textwidth]{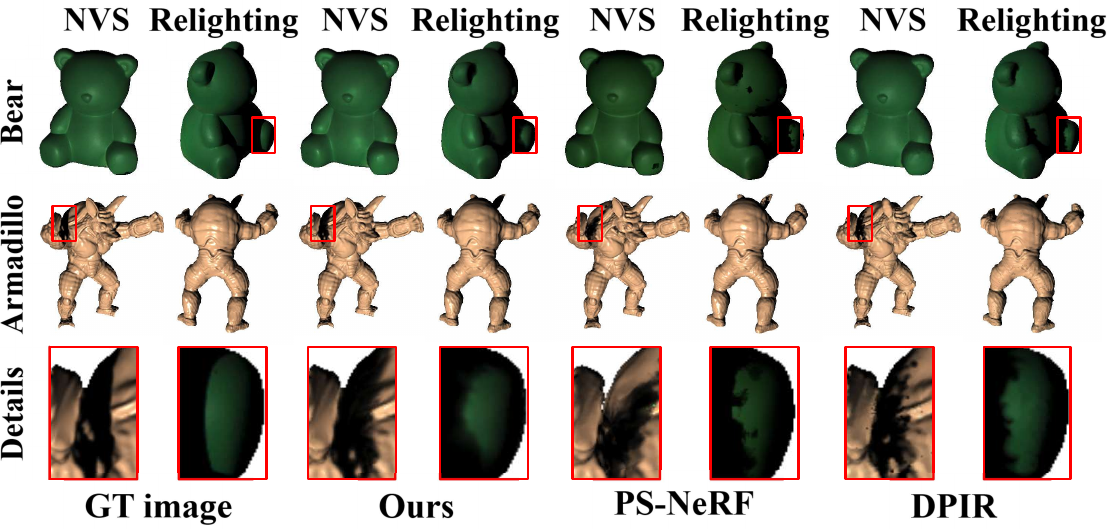} 
\caption{Qualitative comparison of novel view synthesis (NVS) and relighting. The red box displays the details.}
\label{fig2}
\end{figure}

\begin{figure}[htb]
\centering
\includegraphics[width=0.47\textwidth]{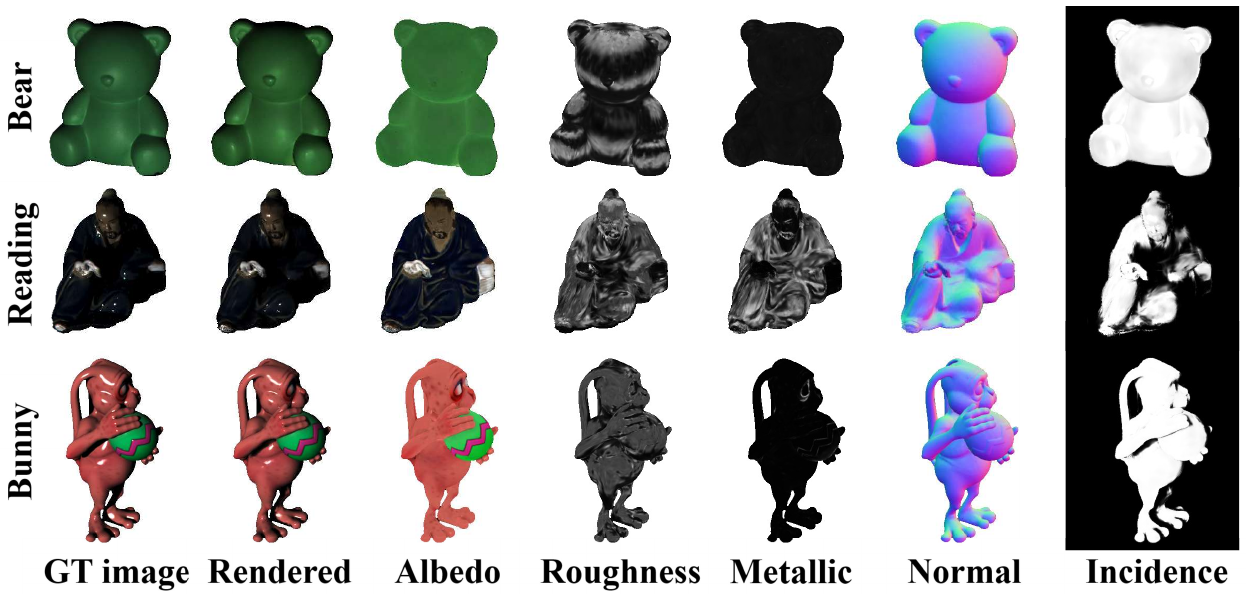} 
\caption{Inverse rendering results on \textbf{Bear}, \textbf{Reading}, and \textbf{Bunny}.}
\label{fig3}
\end{figure}

In Table \ref{table1}, we present the experiment results of NVS and relighting task across DiliGent-MV and PS-NeRF datasets. PS-GS achieves state-of-the-art performance in most objects than previous methods, demonstrating its ability to accurately reconstruct. PS-GS also completes training in a relatively short time of 0.7 hours and low GPU memory usage of 6.5 GB, reflecting its efficiency in MVPS-based inverse rendering. Figure \ref{fig2} illustrates a qualitative comparison against competitors, visualizing rendering images at NVS and relighting task. The results of PS-NeRF have some strange shadows and introduce over-smooth rendered images and relatively low overall intensity. DPIR, which employs point-based splatting and shadow, finds that the edge of its shadow exhibits distinct discrete point-like shadow. In contrast, our Gaussian Splatting based method enables smooth shadow edges, which makes rendering more realistic. We also provide the inverse rendering result in Figure \ref{fig3}, such as the estimated diffuse map, roughness map, metallic map, and incident lighting. PS-GS achieves more compelling reconstruction results. 

\subsubsection{Results for Normal Accuracy} 

\begin{figure}[htb]
\centering
\includegraphics[width=0.47\textwidth]{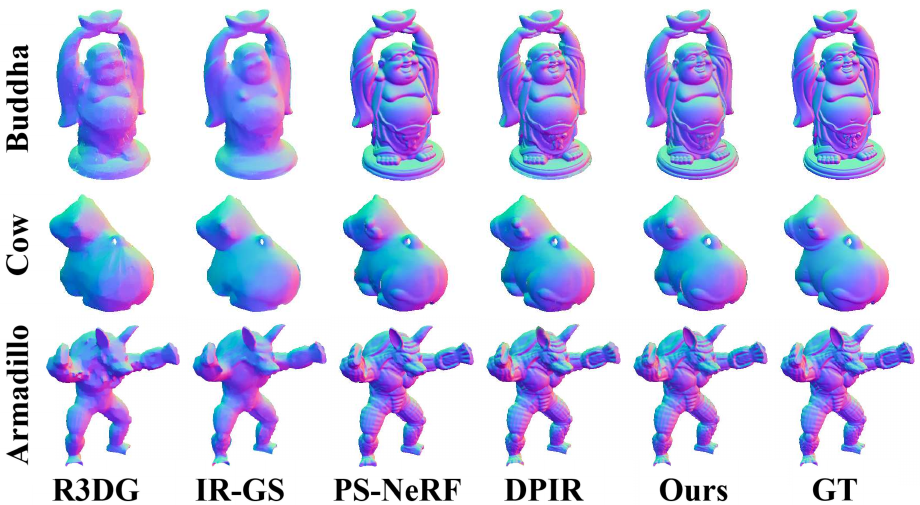} 
\caption{Qualitative comparison on normal accuracy.}
\label{fig4}
\end{figure}

\begin{table*}[t]
\centering
\begin{tabular}{lccccccc|cc}
\toprule
     & Bear & Buddha & Reading & Cow & Pot2 & Armadillo & Bunny & GPU memory$\downarrow$ & Time$\downarrow$\\
\midrule
R3DG & 10.48 & 22.08 & 19.25 & 9.13 & 13.00 & 15.15 & 15.12 & 8 G & 1 h\\
IRGS & 11.48 & 23.38 & 17.80 & 11.54 & 16.16 & 15.51 & 15.60 & 7 G & \textbf{0.7} h\\
PS-NeRF & 5.03 & 12.35 & 9.37 & 5.96 & 7.59 & 5.03 & 5.53 & 9.6 G & \textgreater 22 h\\
DPIR & 4.86 & 11.26 & 8.95 & \textbf{4.36} & 6.50 & 4.41 & 5.53 & 9 G & 2 h\\
Ours & \textbf{3.68} & \textbf{8.92} & \textbf{8.21} & 4.87 & \textbf{6.08} & \textbf{3.53} & \textbf{3.47} & \textbf{6.5} G & \textbf{0.7} h\\
\bottomrule
\end{tabular}

\caption{Qualitative comparison of normal accuracy (quantified by MAE($\downarrow$)) on the objects of DiLiGenT-MV dataset and PS-NeRF synthesis dataset. In this table, the best results are in bold.}
\label{table2}
\end{table*}
Table \ref{table2} shows the quantitative comparison results on estimated normal map between our method with both MVPS-based and Gaussian-based inverse rendering baselines. PS-GS demonstrates superior surface reconstruction compared to other methods, as demonstrated by the high quality of the normal maps. This improvement is largely attributed to the specifically designed geometry modeling and regularization strategies described above. As shown in Figure \ref{fig4}, the normal maps estimated by R3DG and IRGS are too smooth to present the details of the object surface. In contrast, PS-GS overcomes this challenge, achieving accuracy and more details for normal maps. Our approach utilizes multiple illumination images of each view, as opposed to fixed environment illumination, providing more information for reconstruction. Additionally, our method enhances the reasonableness and realism of normal maps efficiently, which is benefited by the advantage of the 2DGS model frame.

\subsection{Ablation Study}
\subsubsection{Regularization for Rendered Normal Maps} 
\begin{figure}[htb]
\centering
\includegraphics[width=0.35\textwidth]{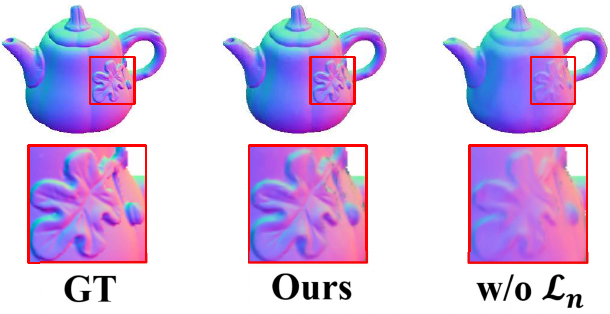} 
\caption{Ablation study on normal regularization. The red box displays the details.}
\label{fig5}
\end{figure}

Benefiting from the surface normals estimated by the UPS method, we regularize the rendered normal maps of the Gaussian model, which can mitigate ambiguity in surface estimation. Figure \ref{fig5} and Table \ref{table3} present the result of our full model and the model without normal regularization. The absence of normal regularization leads to a smooth reconstruction result that lacks details. The full model significantly improves the rendered normal accuracy and recovered surface details.

\subsubsection{Regularization for Incident Lighting} 
\begin{figure}[ht]
\centering
\includegraphics[width=0.4\textwidth]{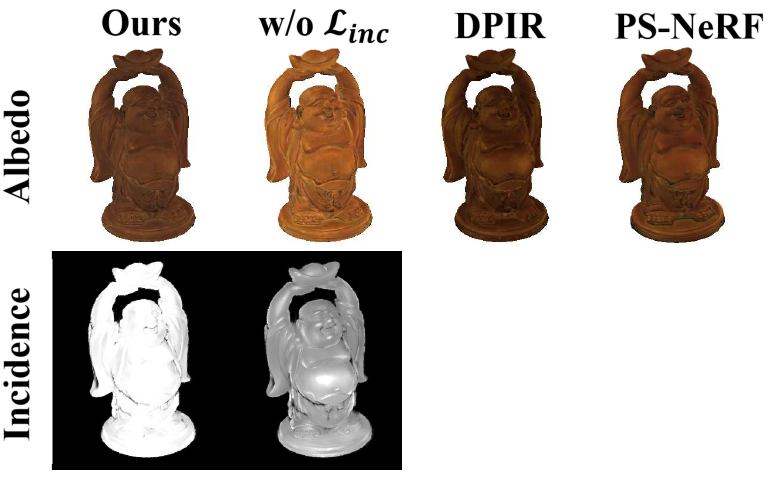} 
\caption{Ablation study on incident lighting regularization. We also provide the diffuse map rendered by DPIR and PS-NeRF for reference.}
\label{fig6}
\end{figure}

We modify the 2D Gaussian ray-tracing technique to be suitable for SDL. We then leverage its visibility result to regularize the MLP-computed incident lighting. As shown in Figure \ref{fig6} and Table \ref{table3}, without using the regularization, the incident lighting is baked into the estimated albedo maps. The regularization helps to estimate relatively accurate intensity for both albedo maps and incident lighting. 
\subsubsection{The Standard 2DGS Model Pretraining} 
\begin{table}[htb]
\setlength{\tabcolsep}{1mm}
\centering
\begin{tabular}{lcccc|c}
\toprule
     & PSNR$\uparrow$ & SSIM$\uparrow$ & \makecell{LPIPS$\downarrow$\\($\times 10^{-2}$)} & MAE$\downarrow$ & Time$\downarrow$\\
\midrule
w/o $\mathcal{L}_n$ & 37.50 & 0.984 & 1.50 & 7.02 & 0.7 h  \\
w/o $\mathcal{L}_{inc}$ & \textbf{37.92} & 0.985 & \textbf{1.37} & 5.64 & 0.7 h \\
w/o Stage \Roman{mycounter1} & 37.05 & 0.984 & 1.51 & 5.69 & 2.0 h \\
10 views & 37.17 & 0.984 & 1.55 & 5.72 & 0.6 h \\
5 views & 34.93 & 0.970 & 2.49 & 8.34 & \textbf{0.5} h\\
Ours & 37.67 & \textbf{0.986} & 1.44 & \textbf{5.57} & 0.7 h\\
\bottomrule
\end{tabular}
\caption{Ablation studies on various components of PS-GS. In this table, the results are averaged among all the objects, and the best results are in bold.}
\label{table3}
\end{table}
The results, as shown in Table \ref{table3}, emphasize the superiority of the standard 2DGS model pretraining over only performing inverse rendering with an equal number of training iterations. Simultaneously recovering the geometry, materials, and lighting only from input images is difficult. The usage of standard 2DGS pretraining process can provide a reliable geometry initialization before the inverse rendering, which offers superiority in both the quality of reconstruction and the speed of shape convergence.
\subsubsection{Experiments on Training Views}
\begin{figure}[htb]
\centering
\includegraphics[width=0.47\textwidth]{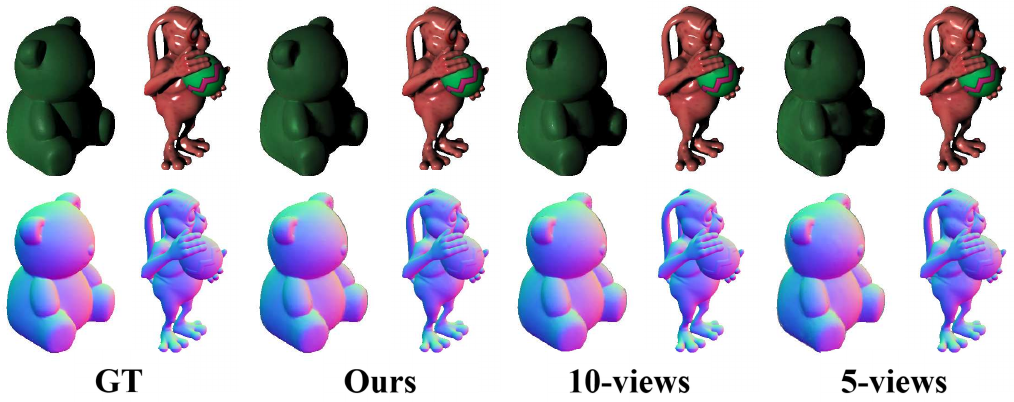} 
\caption{Ablation study on training views.}
\label{fig7}
\end{figure}

Based on the MVPS technique, our method exploits the multi-view and multi-light images in the inverse rendering process, which is able to mitigate the inherent ambiguity of constant illuminated images. Table \ref{table3} and Figure \ref{fig7} show the reconstruction results with varying numbers of training views. Despite the decline of the metric values, our method obtained satisfying reconstruction results even when only a 5-view input is provided, which demonstrates the ability to reconstruct in sparse views.

\subsection{Applications}
\begin{figure}[ht]
\centering
\includegraphics[width=0.38\textwidth]{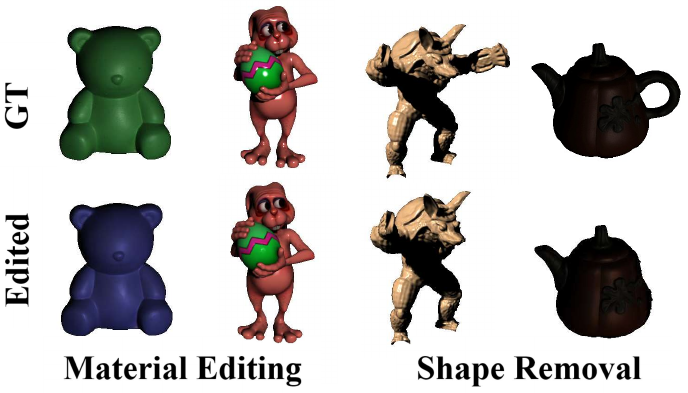} 
\caption{Applications on PS-GS, including shape removal and material editing. For material editing, the first column is the result of replacing the diffuse map with the alternative, and the second column is the result of reducing the intensity of the specular map.}
\label{fig8}
\end{figure}

Owing to the explicit Gaussian primitive representation, materials and geometry editing can be achieved conveniently through PS-GS by editing the parameters of Gaussian primitives. Specifically, PS-GS allows material editing for both diffuse and specular term through replacing them with alternatives or adjusting the intensity of the reconstructed materials. Furthermore, the intuitive shape removal is supported by our method via simply removing the Gaussian primitives from the reconstructed geometry model. The editing results for materials and geometry editing of PS-GS are shown in Figure \ref{fig8}.

\section{Conclusion}
In this paper, we introduce Gaussian splatting for Multi-view Photometric Stereo (PS-GS), an inverse rendering method that integrates 2D Gaussian Splatting and physically-based deferred shading for MVPS. Our method first reconstructs an object with a standard 2D Gaussian model as initialization. It then jointly optimizes geometry, materials, and lighting of the object by multi-view and multi-light images based on the full rendering equation with a lighting prediction network. To mitigate the inherent ambiguity of inverse rendering, we also perform the regularization for the rendered normal maps by the normals obtained from multi-light images via the UPS method. We further propose a regularization on the estimated lighting based on the visibility computed by a modified 2D Gaussian ray-tracing method that is suitable for SDL. Experiments on both synthetic and real datasets demonstrate the superiority of PS-GS over existing methods in terms of quantitative metrics, visual quality, and efficiency. Notably, our method enables high-quality reconstructions even under 5 input views.


\bibliography{ps-gs}
\end{document}